\documentstyle[11pt,aasms4]{article}

\begin{document}

\title{The Visual Orbit of $\iota$ Pegasi}
\author{A.F.~Boden\altaffilmark{1},
	C.D.~Koresko\altaffilmark{3},
	G.T.~van Belle\altaffilmark{1},
	M.M.~Colavita\altaffilmark{1},
	P.J.~Dumont\altaffilmark{1},
	J.~Gubler\altaffilmark{2},
	S.R.~Kulkarni\altaffilmark{3},
	B.F.~Lane\altaffilmark{3},
	D.~Mobley\altaffilmark{1},
	M.~Shao\altaffilmark{1},
	J.K.~Wallace\altaffilmark{1}\\
	(The PTI Collaboration) \\
	and \\
	G.W.~Henry\altaffilmark{4}}
\altaffiltext{1}{Jet Propulsion Laboratory, California Institute of Technology}
\altaffiltext{2}{University of California, San Diego}
\altaffiltext{3}{Palomar Observatory, California Institute of Technology}
\altaffiltext{4}{Center of Excellence in Information Systems, Tennessee State University}
\authoremail{bode@huey.jpl.nasa.gov}

\begin{abstract}
We have determined the visual orbit for the spectroscopic binary
$\iota$~Pegasi with interferometric visibility data obtained by the
Palomar Testbed Interferometer in 1997.  $\iota$~Pegasi is a
double-lined binary system whose minimum masses and spectral typing
suggests the possibility of eclipses.  Our orbital and component
diameter determinations do not favor the eclipse hypothesis: the
limb-to-limb separation of the two components is 0.151 $\pm$ 0.069 mas
at conjunction.  Our conclusion that the $\iota$~Peg system does not
eclipse is supported by high-precision photometric observations.

The physical parameters implied by our visual orbit and the
spectroscopic orbit of Fekel and Tomkin (1983) are in good agreement
with those inferred by other means.  In particular, the orbital
parallax of the system is determined to be 86.9 $\pm$ 1.0 mas, and
masses of the two components are determined to be 1.326 $\pm$ 0.016
M$_{\sun}$ and 0.819 $\pm$ 0.009 M$_{\sun}$ respectively.
\end{abstract}


\section{Introduction}

$\iota$~Pegasi (HR 8430, HD 210027) is a nearby, short-period (10.2 d)
binary system with a F5V primary and a $\sim$ G8V secondary in a
circular orbit.  $\iota$~Peg was first discovered as a single-lined
spectroscopic binary by Campbell (1899), and the first spectroscopic
orbital elements were estimated by Curtis (1904).  Several other
single-line studies were made, notably Petrie and Phibbs (1949) and
Abt and Levy (1976).  In the context of a lithium abundance study,
Herbig (1965) noted that lines from the $\iota$~Peg secondary were
visible at red wavelengths.  Lithium abundances for both the primary
(\cite{Herbig65,Conti66,Duncan81,Lyubimkov91}) and the secondary
(\cite{Fekel83,Lyubimkov91}) indicate the system is very young ($\sim$
8 $\times$ 10$^7$ yr, \cite{Fekel83}, 1.7 $\pm$ 0.8 $\times$ 10$^8$
yr, \cite{Lyubimkov91}) and both components are close to the zero-age
main sequence.  Both components of $\iota$~Peg are also believed to
have solar-type abundances (\cite{Lyubimkov91}).

Following Herbig's implicit suggestion, Fekel and Tomkin (1983,
hereafter FT) made radial velocity measurements of both $\iota$~Peg
components at 643 nm, and computed a definitive spectroscopic orbit
and inferred a probable G8V spectral classification for the secondary.
FT's orbit was noteworthy as it indicated that the minimum masses for
the two components were very near the model values for the spectral
types, suggesting a ``reasonable prospect'' for eclipses in the system
(FT).  Subsequent photometric monitoring by automated photometry
projects in Arizona, at Palomar Observatory, and in Pasadena failed to
show any evidence for eclipses (see \S \ref{sec:eclipses}).  FT also
questioned synchronous rotation of the secondary.  However, Gray
(1984), from somewhat higher resolution spectroscopic data, argued
that both components are in synchronous rotation.

Herein we report a determination of the $\iota$~Peg visual orbit from
near-infrared, long-baseline interferometric visibility measurements
taken with the Palomar Testbed Interferometer.  PTI is a 110-m K-band
(2 - 2.4 $\mu$m) interferometer located at Palomar Observatory, and
described in detail elsewhere (\cite{Colavita94,Colavita98a}).  The
minimum PTI fringe spacing is roughly 4 mas at the sky position of
$\iota$~Peg, allowing us to resolve this binary system.  The
procedures we have used to determine $\iota$~Peg's visual orbit are
similar to other visual orbits determined for spectroscopic binaries
using the Mark III Interferometer at Mt.~Wilson
(\cite{Pan90,Armstrong92a,Armstrong92b,Pan92,Hummel93,Pan93,Hummel94,Hummel95}),
and the NPOI Interferometer at Anderson Mesa, AZ (\cite{Hummel98}).
The analogy between $\iota$~Peg and the short-period, small angular
scale binaries studied in Hummel et al.~(1995) and Hummel et
al.~(1998) is especially apt.

\section{Observations}
Pan attempted to determine a visual orbit for $\iota$~Peg using the
Mark III interferometer at Mt.~Wilson, but the significant brightness
difference in the two components at 800 nm made the observations
difficult (\cite{Pan97}).  The apparent contrast ratio in the
$\iota$~Peg system decreases in the K-band, allowing a reliable orbit
determination with PTI observations.

The observable used for these observations is the fringe contrast or
{\em visibility} (squared) of an observed brightness distribution on
the sky.  Normalized in the interval [0,1], a single star exhibits
visibility modulus given in a uniform disk model by:
\begin{equation}
V =
\frac{2 \; J_{1}(\pi B \theta / \lambda)}{\pi B \theta / \lambda}
\label{eq:V_single}
\end{equation}
where $J_{1}$ is the first-order Bessel function, $B$ is the projected
baseline vector magnitude at the star position, $\theta$ is the
apparent angular diameter of the star, and $\lambda$ is the
center-band wavelength of the interferometric observation.  (We
consider corrections to the uniform disk model from limb darkening in
\S \ref{sec:physics}.)  The expected squared visibility in a narrow
pass-band for a binary star such as $\iota$~Peg is given by:
\begin{equation}
V^{2}_{nb}(\lambda)
 = \frac{V_{1}^2 + V_{2}^2 \; r^2 + 2 \; V_{1} \; V_{2} \; r \;
	  \cos(\frac{2 \pi}{\lambda} \; {\bf {B}} \cdot {\bf {s}})}
	{(1 + r)^2}
\label{eq:V2_double}
\end{equation}
where $V_{1}$ and $V_{2}$ are the visibility moduli for the two stars
alone as given by Eq.~\ref{eq:V_single}, $r$ is the apparent
brightness ratio between the primary and companion, ${\bf {B}}$ is the
projected baseline vector at the system sky position, and ${\bf {s}}$
is the primary-secondary angular separation vector on the plane of the
sky (\cite{Pan90,Hummel95}).  The $V^2$ observables used in our
$\iota$ Peg study are both narrow-band $V^2$ from seven individual
spectral channels (\cite{Colavita98a}), and a synthetic wide-band
$V^2$, given by an incoherent SNR-weighted average $V^2$ of the
narrow-band channels in the PTI spectrometer (\cite{Colavita98b}).  In
this model the expected wide-band $V^2$ observable is approximately
given by an average of the narrow-band formula over the finite
pass-band of the spectrometer:
\begin{equation}
V^{2}_{wb} = \frac{1}{n}\sum_{i}^{n} V^{2}_{nb-i}(\lambda_i)
\label{eq:V2_doubleWB}
\end{equation}
where the sum runs over the n = 7 channels with wavelengths
$\lambda_i$ covering the K-band (2 - 2.4 $\mu$m) of the PTI
spectrometer in its 1997 configuration.  Separate calibrations and
hypothesis fits to the narrow-band and synthetic wide-band $V^2$
datasets yield statistically consistent results, with the synthetic
wide-band data exhibiting superior fit performance.  Consequently we
will present only the results from the synthetic wide-band data.

$\iota$~Peg was observed by PTI on 24 nights between 2 July and 8 Sept
1997.  In each night $\iota$~Peg was observed in conjunction with
calibration objects multiple times during the night.  Each observation
(``scan'') was from 120 -- 130 seconds in duration.  For each scan we
computed a mean $V^2$ value through methods described in Colavita
(1999b).  We assumed the measured rms in the internal scatter to be
the error in $V^2$.  For the purposes of this analysis we have
restricted our attention to four calibration objects, two primary
calibrators within 5$^{\circ}$ of $\iota$~Peg (HD 211006 and HD
211432), and two ancillary calibrators within 15$^{\circ}$ of
$\iota$~Peg (HD 215510 and HD 217014 -- 51~Pegasi).  The suitability
of 51~Peg (a known radial velocity variable) as a calibrator at PTI is
addressed in Boden et al.~(1998b).  Table \ref{tab:calibrators}
summarizes the relevant parameters on the calibration objects used in
this study.  In particular we have estimated our calibrator diameters
based on a model diameter on 51~Peg of 0.72 $\pm$ 0.06 mas implied by
a linear diameter of 1.2 $\pm$ 0.1 R$_{\sun}$ (adopted by
\cite{Marcy97}) and a parallax of 65.1 $\pm$ 0.76 mas from Hipparcos
(\cite{HIP97,Perryman97}).

The calibration of $\iota$~Peg $V^2$ data is performed by estimating
the interferometer system visibility ($V^{2}_{sys}$) using calibration
sources with model angular diameters, and then normalizing the raw
$\iota$~Peg visibility by $V^{2}_{sys}$ to estimate the $V^2$ measured
by an ideal interferometer at that epoch
(\cite{Mozurkewich91,Boden98a}).  We calibrated the $\iota$~Peg $V^2$
data in two different ways: (1) with respect to the two primary
calibration objects, resulting in our primary dataset containing 112
calibrated observations over 17 nights, and (2) an unbiased average of
the primary and ancillary calibrators, resulting in our secondary
dataset containing 151 observations over 24 nights.  The motivation
for constructing these two datasets, which are clearly not
independent, is that the determination of the orbital solution and
component diameters is sensitive to calibration uncertainties.
Comparison of the solutions derived from the two datasets allow us to
quantitatively assess this uncertainty.

\begin{table}[t]
\begin{center}
\begin{small}
\begin{tabular}{|c|c|c|c|c|}
\hline
Object    & Spectral & Star        & Sky Separation   & Diam.~WRT     \\
Name      & Type     & Magnitude   & From $\iota$~Peg & Model 51 Peg  \\
\hline
HD 211006 & K2III    & 5.9 V/3.4 K & 3.6$^{\circ}$ & 1.06 $\pm$ 0.05   \\
HD 211432 & G9III    & 6.4 V/3.7 K & 3.2$^{\circ}$ & 0.70 $\pm$ 0.05   \\
\hline \hline
HD 215510 & G6III    & 6.3 V/3.9 K & 11$^{\circ}$  & 0.85 $\pm$ 0.06   \\
HD 217014 & G2.5V    & 5.9 V/4.0 K & 12$^{\circ}$  & (0.72 $\pm$ 0.06)  \\
\hline
\end{tabular}
\caption{1997 PTI $\iota$~Peg Calibration Objects Considered in our
Analysis.  The relevant parameters for our four calibration objects
are summarized.  The apparent diameter values are determined by a fit
to our $V^2$ data calibrated with respect to a model diameter for HD
217014 (51 Peg) of 0.72 $\pm$ 0.06 mas
(\cite{Marcy97,HIP97}).
\label{tab:calibrators}}
\end{small}
\end{center}
\end{table}

\section{Orbit Determination}

The estimation of the $\iota$~Peg visual orbit is made by fitting a
Keplerian orbit model with visibilities predicted by
Eqs.~\ref{eq:V2_double} and \ref{eq:V2_doubleWB} directly to the
calibrated (narrow-band and synthetic wide-band) $V^2$ data on
$\iota$~Peg (see \cite{Armstrong92b,Hummel93,Hummel95}).  The fit is
non-linear in the Keplerian orbital elements, and is therefore
performed by non-linear least-squares methods (i.e.~the
Marquardt-Levenberg method, \cite{Press92}).  As such, this fitting
procedure takes an initial estimate of the orbital elements and other
parameters (e.g. component angular diameters, brightness ratio), and
refines the model into a new parameter set which best fits the data.
However, the chi-squared surface has many local minima in addition to
the global minimum corresponding to the true orbit.  Because
Marquardt-Levenberg strictly follows a downhill path in the $\chi^2$
manifold, it is necessary to thoroughly survey the space of possible
binary parameters to distinguish between local minima and the true
global minimum.  In the case of $\iota$~Peg the parameter space is
significantly narrowed by the high-quality spectroscopic orbit and
inclination constraint near 90$^\circ$ (FT).  Furthermore, the
Hipparcos distance determination sets the rough scale of the
semi-major axis (\cite{HIP97}).

In addition, as the $V^2$ observable for the binary
(Eqs.~\ref{eq:V2_double} and \ref{eq:V2_doubleWB}) is invariant under
a rotation of 180$^{\circ}$, we cannot differentiate between an
apparent primary/secondary relative orientation and its mirror image
on the sky.  In order to follow the FT convention for T$_0$ at primary
radial velocity maximum, in our analysis of $\iota$~Peg we have
defined T$_0$ to be at a component separation extremum, yielding an
extremum in component radial velocities for the circular orbit.  We
have additionally required our fit T$_0$ to be within half a period of
the projected FT determination to differentiate between primary radial
velocity maximum and minimum.  Even with our determination of T$_0$ so
defined there remains a 180$^{\circ}$ ambiguity in our determination
of the longitude of the ascending node, $\Omega$.

We used a preliminary orbital solution computed by Pan (1996) by
separation vector techniques (see \cite{Pan90} for a discussion of the
method), and refined it into the best-fit orbit shown here.  We
further conducted an exhaustive search of the binary parameter space
that resulted in the same best-fit orbit, which is in fact the global
minimum in the $\chi^2$ manifold.

\begin{figure}
\epsscale{0.7}
\plotone{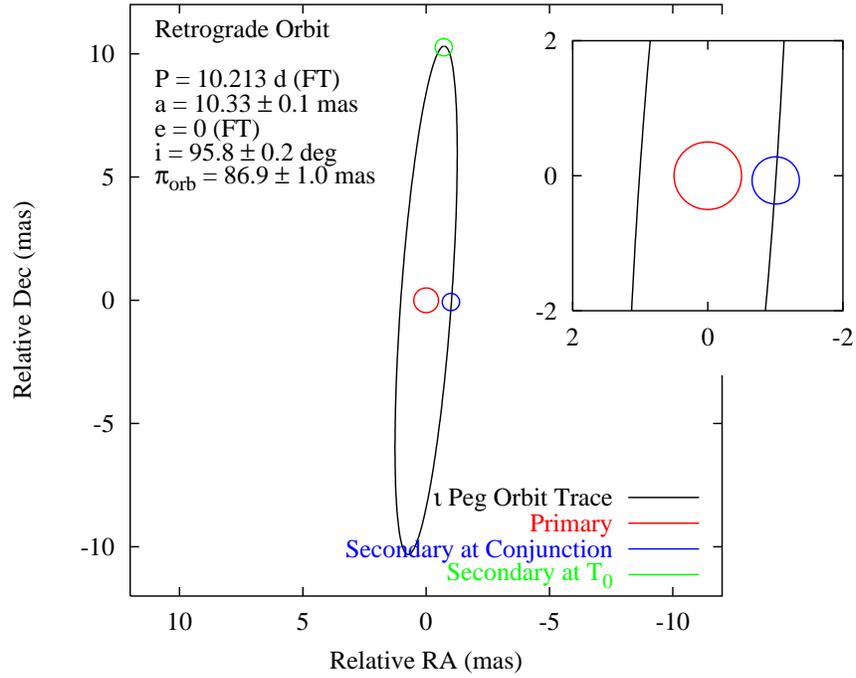}
\caption{Visual Orbit of $\iota$~Pegasi.  The relative visual orbit of
$\iota$~Peg is depicted, with the primary and secondary rendered at
T$_0$ (maximum primary radial velocity) and apparent conjunction.  The
inset shows a closeup of the system at apparent conjunction.  By our
model the $\iota$ Peg orbit is nearly, but not quite eclipsing, being
approximately 0.87$^\circ$ in inclination from apparent grazing
eclipses.
\label{fig:iPg_orbit}}
\end{figure}

Figure \ref{fig:iPg_orbit} depicts the apparent relative orbit of the
$\iota$~Peg system.  Most striking is the observation that the
circular orbit of the system (see below) is very nearly eclipsing.
From our primary dataset we find a best fit orbital inclination of
95.67 $\pm$ 0.21 degrees.  With model angular diameters of 1.0 and 0.7
mas for the primary and secondary components respectively (\S
\ref{sec:physics}), and an apparent semi-major axis of 10.33 $\pm$
0.10 mas, this inclination is about 0.87$^{\circ}$ from apparent
limb-to-limb contact.  This is consistent with the lack of photometric
evidence for eclipses despite several photometry campaigns on the
$\iota$~Peg system (\S \ref{sec:eclipses}).

\begin{table}
\dummytable\label{tab:dataTable}
\end{table}

\begin{figure}
\epsscale{0.8}
\plotone{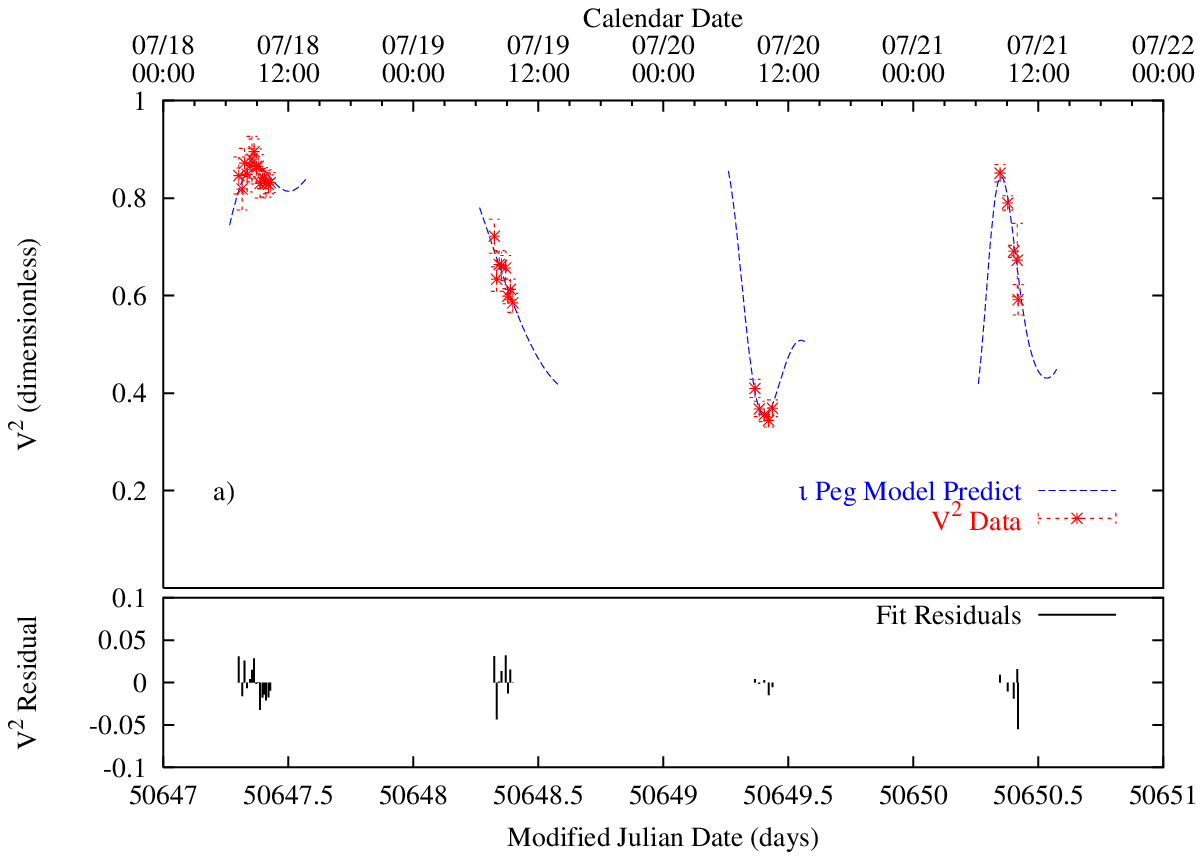}\\
\plotone{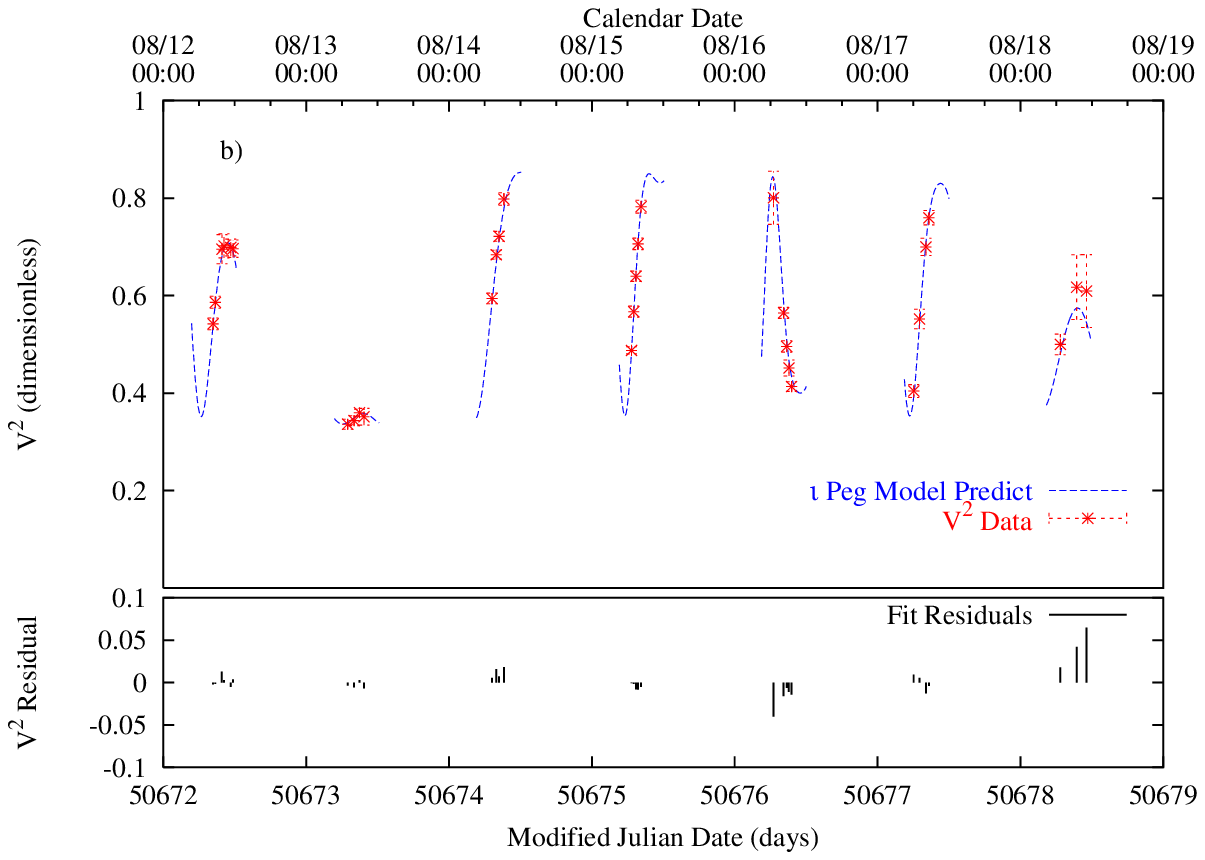}
\caption{$V^2$ Fit of $\iota$~Pegasi.  a) Four consecutive nights (18
-- 21 July 1997) of calibrated $V^2$ data on $\iota$~Peg, and $V^2$
predictions from the best-fit model for the system.  In the lower
frame we give $V^2$ residuals between the calibrated data and best-fit
model.  b) An additional seven consecutive nights (12 -- 18 August 1997) of
data on $\iota$~Peg, with model predicts and fit residuals.  The model
is in good agreement with the calibrated data, with a $\chi^2$/DOF of
0.75 and an average absolute $V^2$ residual of 0.014.
\label{fig:iPg_fit}}
\end{figure}

Table \ref{tab:dataTable} lists the complete set of $V^2$ measurements
in the primary dataset and the prediction based on the best-fit orbit
model for $\iota$~Peg.  Figure \ref{fig:iPg_fit} shows two graphical
comparisons between our $V^2$ data on $\iota$~Peg and the best-fit
model predictions.  Figure \ref{fig:iPg_fit}a gives four consecutive
nights of PTI $V^2$ data from our primary dataset on $\iota$~Peg (18
-- 21 July 1997), and $V^2$ predictions based on the best-fit model
for the system.  Figure \ref{fig:iPg_fit}b gives an additional seven
consecutive nights (12 -- 18 August 1997) with the same quantities
plotted.  These are the two longest consecutive-night sequences in our
data set.  The model predictions are seen to be in excellent absolute
and statistical agreement with the observed data, with a primary
dataset average absolute $V^2$ deviation of 0.014, and a $\chi^2$ per
Degree of Freedom (DOF) of 0.75.

\begin{figure}
\epsscale{0.8}
\plotone{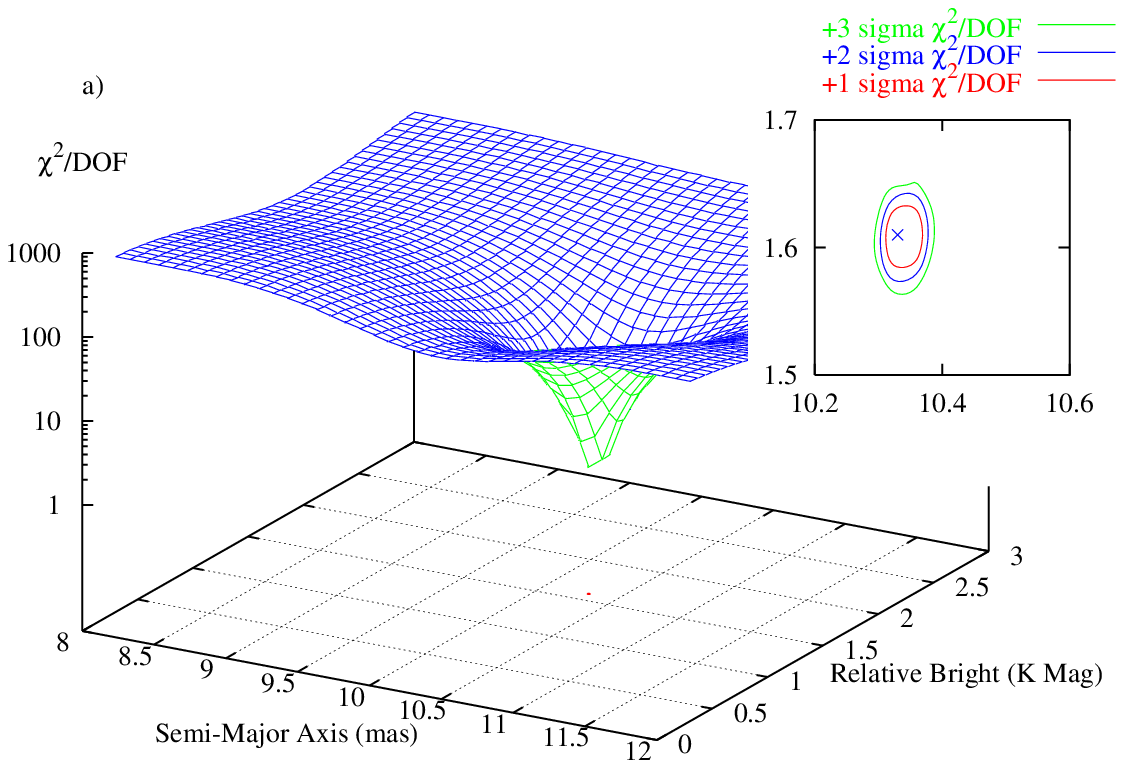}\\
\plotone{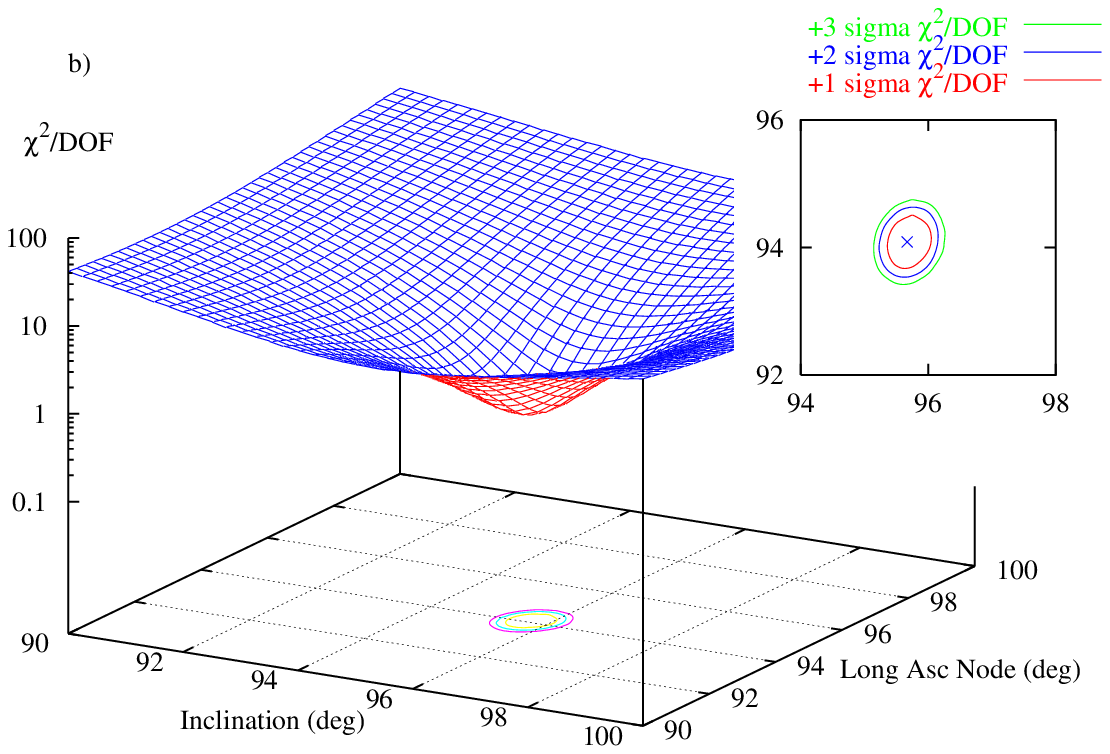}
\caption{$\chi^2$/DOF Fit Surfaces for $\iota$~Pegasi Primary Dataset.
a) $\chi^2$/DOF surface in the subspace of orbit semi-major axis and
relative component brightness.  Inset is a closeup of a contour plot
surface indicating location of the best-fit parameter values, and
contours at +1, +2, and +3 of $\chi^2$/DOF significance.  b)
$\chi^2$/DOF surface in the subspace of orbital inclination and
longitude of the ascending node, with inset giving surface contour
closeup.
\label{fig:iPg_surf}}
\end{figure}

Figure \ref{fig:iPg_surf} gives two examples of the $\chi^2$ fit
projected into orbital parameter subspaces.  Figure
\ref{fig:iPg_surf}a shows a surface of $\chi^2$/DOF projected into the
subspace of orbit semi-major axis and relative component brightness,
with all other parameters held to their best-fit values.  Inset is a
closeup of a contour plot of the $\chi^2$/DOF surface indicating
location of the best-fit parameter values, and contours at +1, +2, and
+3 of $\chi^2$/DOF significance.  Figure \ref{fig:iPg_surf}b gives the
$\chi^2$/DOF surface in the subspace of orbital inclination and
longitude of the ascending node.  Again, the inset gives best-fit
parameter values, and contours at +1, +2, and +3 of $\chi^2$/DOF
significance.  All indications are that the best-fit model for the
$\iota$~Peg system is in excellent agreement with our $V^2$ data, and
that data uniquely constrain the parameters of the visual orbit.

Spectroscopic (from FT) and visual orbital parameters of the $\iota$
Peg system are summarized in Table \ref{tab:orbit}.  We present the
results for our primary and secondary datasets separately.  For the
parameters we have estimated from our interferometric data we quote a
total one-sigma error in the parameter estimates, and the one-sigma
errors in the parameter estimates from statistical (measurement
uncertainty) and systematic error sources.  In our analysis the
dominant forms of systematic error are: (1) uncertainties in the
calibrator angular diameters (Table \ref{tab:calibrators}); (2) the
uncertainty in our center-band operating wavelength ($\lambda_0
\approx$ 2.2 $\mu$m), which we have taken to be 20 nm ($\sim$1\%); (3)
the geometrical uncertainty in our interferometric baseline ( $<$
0.01\%); and (4) uncertainties in orbital parameters we have
constrained in our fitting procedure (e.g. period, eccentricity).
Different parameters are affected differently by these error sources;
our estimated uncertainty in the $\iota$~Peg orbital inclination is
dominated by measurement uncertainty, while the uncertainty in the
angular semi-major axis is dominated by uncertainty in the wavelength
scale.  Conversely, we have assumed that all the uncertainty quoted by
FT in the $\iota$~Peg spectroscopic parameters is statistical.
Finally, we have listed the level of statistical agreement in the
visual orbit parameters in our two solutions (the absolute residual
between the two estimates divided by the RSS of their statistical
errors).  The two solutions are in good statistical agreement, giving
us confidence we have properly characterized our calibration
uncertainties.

Particularly remarkable is the agreement between T$_{0}$ (quoted as
the epoch of maximum primary radial velocity for the $\iota$~Peg
circular orbit) and period as determined by FT, and T$_{0}$ as
determined in our primary dataset, separated from the FT determination
by 523 cycles.  FT quote an $\iota$~Peg period accurate to roughly 1
part in 10$^{6}$, resulting in a propagated uncertainty in T$_{0}$ at
the epoch of our observations of 7 $\times$ 10$^{-3}$ days.  This
FT-extrapolated T$_{0}$ differs from our 1997 T$_{0}$ determination by
8 $\times$ 10$^{-4}$ days, an agreement of roughly 0.1 sigma.  A
similar comparison with the secondary dataset solution is less
spectacular, an agreement at 0.7 sigma.  Clearly the extraordinary
quoted accuracy of the $\iota$~Peg period determination by FT (made by
combining their 1977 -- 1982 data with spectroscopy from the mid-30s
-- \cite{Petrie49}) seems well justified compared to our visual orbit.
Consequently we have assumed the FT value for the $\iota$~Peg period.

Following FT we have assumed a circular orbit for the system.  Fitting
our primary dataset for an eccentricity in the system yields an
estimate of 1.5 $\times$ 10$^{-3}$ $\pm$ 1.3 $\times$ 10$^{-3}$.  The
assumption of a circular orbit seems well justified.

\begin{table}
\begin{center}
\begin{small}
\begin{tabular}{|c|c||c|c|c|}
\hline
Orbital			&   FT    	& \multicolumn{3}{c|}{PTI 1997} \\
\cline{3-5}
Parameter       	& 1983    	& Primary Dataset	& Secondary Dataset & Stat Agr \\
\hline \hline
Period (d)      	& 10.213033     &  10.213033  		& 10.213033	 & 		\\
                	& $\pm$ 1.3 $\times$ 10$^{-5}$ & (assumed) & (assumed)	 &              \\
T$_{0}$ (HJD)   	& 2445320.1423 	& 2450661.5578 		& 2450661.5634	 &  1.26	\\
                	&              	& $\pm$ 3.6 (3.3/1.5) $\times$ 10$^{-3}$ & $\pm$ 3.3 (3.0/1.5) $\times$ 10$^{-3}$ & \\
$e$               	& 0 (assumed) 	& 0 (assumed) 		& 0 (assumed)    & \\
K$_A$ (km s$^{-1}$)   	& 48.1 $\pm$ 0.2 &  			& 		 & \\
K$_B$ (km s$^{-1}$)   	& 77.9 $\pm$ 0.3 &  			& 		 & \\
\hline
$i$ (deg) 		&	& 95.67 $\pm$ 0.22 (0.22/0.03) 	& 96.03 $\pm$ 0.20 (0.20/0.03)  & 1.21 \\
$\Omega$ (deg)		&       & 94.09 $\pm$ 0.23 (0.22/0.05)  & 94.03 $\pm$ 0.25 (0.24/0.05) & 0.03 \\
$a$ (mas) 		&       & 10.33 $\pm$ 0.10 (0.02/0.10)	& 10.32 $\pm$ 0.11 (0.02/0.11)  & 0.35 \\
$\Delta$ K (mag) 	&    	& 1.610 $\pm$ 0.021  		& 1.610 $\pm$ 0.021  		& 0.23 \\
			&       & (0.007/0.020)			& (0.007/0.020) 		& \\
\hline
$\chi^2$/DOF		&	& 0.75				& 1.0				& \\
$\overline{|R_{V^2}|}$ 	&       & 0.014				& 0.016				& \\
N$_{scans}$		&       & 112				& 151				& \\
\hline
\end{tabular}
\end{small}
\caption{Orbital Parameters for $\iota$~Peg.  Summarized here are the
apparent orbital parameters for the $\iota$~Peg system as determined
by FT, and our PTI primary and secondary datasets.  For parameters
estimated from our PTI observations we separately quote one sigma
errors from both statistical and systematic sources (listed as
$\sigma_{stat}$/$\sigma_{sys}$), and the total error as the sum of the
two in quadrature.  We have also included the level of statistical
agreement between visual orbit parameters from our two solutions; the
parameters estimated separately from the primary and secondary
datasets are in good agreement in relation to the statistical
component of their error estimates.  We have quoted the longitude of
the ascending node parameter ($\Omega$) as the angle between local
East and the orbital line of nodes (and the relative position of the
secondary at T$_0$), measured positive in the direction of local
North.  Due to the degeneracy in our $V^2$ observable there is a
180$^\circ$ ambiguity in $\Omega$.  Finally, the fit $\chi^2$/DOF and
mean absolute $V^2$ residual ($\overline{|R_{V^2}|}$) is listed for
both solutions.
\label{tab:orbit}}
\end{center}
\end{table}

\section{Physical Parameters}
\label{sec:physics}
Physical parameters derived from the $\iota$~Peg primary dataset
visual orbit and the FT spectroscopic orbit are summarized in Table
\ref{tab:physics}.  We use the primary dataset solution because it is
the most free from possible sky position-dependent systematic effects
(as the secondary dataset includes the ancillary calibrators), but we
note the two orbital solutions yield statistically consistent results.
Notable among the physical parameters for the system is the
high-precision determination of the component masses for the system, a
virtue of the precision of the FT radial velocities on both components
and the high inclination of the orbit.  We estimate the masses of the
F5V primary and putative G8V secondary components as 1.326 $\pm$ 0.016
M$_{\sun}$ and 0.819 $\pm$ 0.009 M$_{\sun}$ respectively.  Our mass
values agree well with mass estimates of 1.33 $\pm$ 0.08 M$_{\sun}$ and
0.9 $\pm$ 0.2 M$_{\sun}$ respectively made by Lyubimkov et al.~(1991)
based on evolutionary models and spectroscopic measurements of component
effective temperatures and surface gravities.

The Hipparcos catalog lists the parallax of $\iota$~Peg as 85.06 $\pm$
0.71 mas (\cite{HIP97}).  The distance determination to
$\iota$ Peg based on the FT radial velocities and our apparent
semi-major axis and inclination is 11.51 $\pm$ 0.13 pc, corresponding
to an orbital parallax of 86.91 $\pm$ 1.0 mas, consistent with the
Hipparcos result at roughly 2\% and 1.5 sigma.

FT list main-sequence model linear diameters for the two $\iota$~Peg
components as 1.3 and 0.9 R$_{\sun}$ respectively (FT).  At a distance
of approximately 11.5 pc this corresponds to apparent angular
diameters of 1.0 and 0.7 mas for the primary and secondary components
respectively.  We have fit for the uniform-disk angular diameter for
both components as a part of the orbit estimation, and find best fit
apparent diameters of 0.98 $\pm$ 0.05 and 0.70 $\pm$ 0.10 mas.
Because we have limited spatial frequency coverage in our data,
following Mozurkewich et al.~(1991) and Quirrenbach et al.~(1996) we
have estimated the limb-darkened diameters of the components from a
correction to the uniform-disk diameter based on the solar
limb-darkening at 2 $\mu$m given by Allen (1982).  The limb-darkened
diameters for the primary and secondary components are 1.0 $\pm$ 0.05
and 0.71 $\pm$ 0.10 mas respectively.  For both the primary and
secondary components our fits for apparent diameter are in good
agreement with main-sequence model diameters.

The observed K-magnitude of the $\iota$~Peg system (2.623 $\pm$ 0.016
-- \cite{Carrasco91}, 2.656 $\pm$ 0.002 -- \cite{Bouchet91}) and our
estimates of the distance and relative K-photometry (Table
\ref{tab:orbit}) of the system allows the determination of the
absolute magnitude of both components separately.  Using the Bouchet
et al.~(1991) K-photometry we obtain M$_{K}$ values of 2.574 $\pm$
0.025 and 4.182 $\pm$ 0.030 for the primary and secondary components
respectively.  Both of these M$_{K}$ values are consistent (within
quoted scatter) to the empirical mass-luminosity relation for nearby
low-mass, main-sequence stars given by Henry \& McCarthy (1992, 1993).
In particular, our M$_{K}$ value for the primary is 0.010 mag brighter
than the mass-luminosity prediction (\cite{Henry92}), while the 4.18
M$_{K}$ value for the secondary is roughly 0.28 magnitudes dimmer than
the prediction (\cite{Henry93}).  Both values are well within the
quoted scatter of the mass-luminosity models.  A second check on the
absolute K-magnitude estimates can be extracted from the model
calculations of Bertelli et al.~(1994), who predict absolute
K-magnitudes of 2.616 $\pm$ 0.048 and 4.254 $\pm$ 0.039 for our
estimated primary and secondary masses respectively for main-sequence
stars with solar-type abundances at an age of 1.7 $\pm$ 0.8 $\times$
10$^8$ yr (\cite{Lyubimkov91}).

\begin{table}
\begin{center}
\begin{small}
\begin{tabular}{|c|c|c|}
\hline
Physical	 & Primary           & Secondary \\
Parameter        & Component         & Component \\
\hline \hline
a (10$^{-2}$ AU) & 4.54 $\pm$ 0.03 (0.03/0.0002)     & 7.35 $\pm$ 0.03 (0.03/0.0003)  \\
Mass (M$_{\sun}$)& 1.326 $\pm$ 0.016 (0.016/0.0001)   & 0.819 $\pm$ 0.009 (0.009/0.0001)  \\
Sp Type (FT)     & F5V               & G8V              \\
Model Diameter (mas) & 1.0	     & 0.7		\\
UD Fit Diameter (mas)& 0.98 $\pm$ 0.05 (0.01/0.05) & 0.70 $\pm$ 0.10 (0.03/0.10)   \\
LD Fit Diameter (mas)& 1.0 $\pm$ 0.05 (0.01/0.05) & 0.71 $\pm$ 0.10 (0.03/0.10)   \\
\cline{2-3}
System Distance (pc) & \multicolumn{2}{c|}{11.51 $\pm$ 0.13 (0.05/0.12)} \\
$\pi_{orb}$ (mas)    & \multicolumn{2}{c|}{86.91 $\pm$ 1.0 (0.34/0.94)} \\
\cline{2-3}
M$_K$ (mag)      & 2.574 $\pm$ 0.025 (0.010/0.024)   & 4.182 $\pm$ 0.030 (0.019/0.028)  \\
\hline
\end{tabular}
\end{small}
\caption{Physical Parameters for $\iota$~Peg.  Summarized here are the
physical parameters for the $\iota$~Peg system as derived from the
orbital parameters in Table \ref{tab:orbit}.  As for our PTI-derived
orbital parameters we have quoted both total error and separate
contributions from statistical and systematic sources (given as
$\sigma_{stat}$/$\sigma_{sys}$).
\label{tab:physics}}
\end{center}
\end{table}

\section{Eclipse Search}
\label{sec:eclipses}
A critical test of our visual orbit model is a high-precision
photometric search for eclipses in $\iota$~Peg.  Combined with our
visual orbit (Table \ref{tab:orbit}), our measured diameters (Table
\ref{tab:physics}) imply an apparent limb-to-limb separation at
conjunction of 0.151 $\pm$ 0.069 mas (using our limb-darkened diameter
estimates).  Our visual orbit and fit diameters do not favor the FT
conjecture of possible eclipses in the $\iota$~Peg system.
Conversely, were the inclination of the orbit near 90$^\circ$, there
would be significant primary eclipses with a duration of a few hours
(6.8 hr for $i$ = 90$^\circ$ -- FT), and as large as 0.6 mag in
V-band.

Several individuals have searched for signs of eclipses in the
$\iota$~Peg system.  In 1997 both Van Buren with the 60'' telescope at
Palomar (1997) and one of us (C.D.K.) at the Robinson Rooftop
Observatory at Caltech in Pasadena (\cite{Koresko97}) searched for
eclipses during primary and secondary eclipse opportunities
respectively.  Both searches resulted in non-detections at about the
0.1 mag levels.


More comprehensive and sensitive than the Southern California searches
has been the program conducted by the Automated Astronomy Group at
Tennessee State University.  $\iota$~Peg was observed photometrically
in 1984 with the Phoenix-10 automatic photoelectric telescope (APT) in
Phoenix, AZ, and again in 1997-98 with the Vanderbilt/Tennessee State
16-inch APT at Fairborn Observatory near Washington Camp, AZ, in order
to search for possible eclipses suggested by FT.  Both telescopes
observed $\iota$~Peg once per night through a Johnson V filter with
respect to the comparison star HR 8441 (HD 210210, F1 IV) in the
sequence C,V,C,V,C,V,C, where C is the comparison star and V is
$\iota$~Peg.  Three differential magnitudes (in the sense V-C) were
computed from each nightly sequence, corrected for differential
extinction, and transformed to the Johnson system.  The three
differential magnitudes from each sequence were then averaged together
and treated as single observations thereafter.  Because of the lack of
accurate standardization in the Phoenix-10 data set, a -0.027 mag
correction was added to each observation to bring those data in line
with the 16-inch observations.  The observations are summarized in
Table \ref{tab:photometry}.  Column 4 gives the standard deviation of
a single nightly observation from the mean of the entire data set and
represents a measure of the precision of the observations.  Further
details on the telescopes, data acquisition, reductions, and quality
control can be found in Young et al.~(1991) and Henry (1995a,b).

\begin{table}
\begin{center}
\begin{small}
\begin{tabular}{|c|c|c|c|}
\hline
APT     &  JD Range        &  \# Obs.  &  Std.~Dev. \\
        &  (+2400000)      &           &   (mag)  \\
\hline
10-inch & 45703 -- 46065   &   78      & 0.0109 \\
16-inch & 50718 -- 50829   &   66      & 0.0032 \\
\hline
\end{tabular}
\end{small}
\caption{Summary of APT Photometry on $\iota$~Peg.
\label{tab:photometry}}
\end{center}
\end{table}

\begin{figure}
\epsscale{0.8}
\plotone{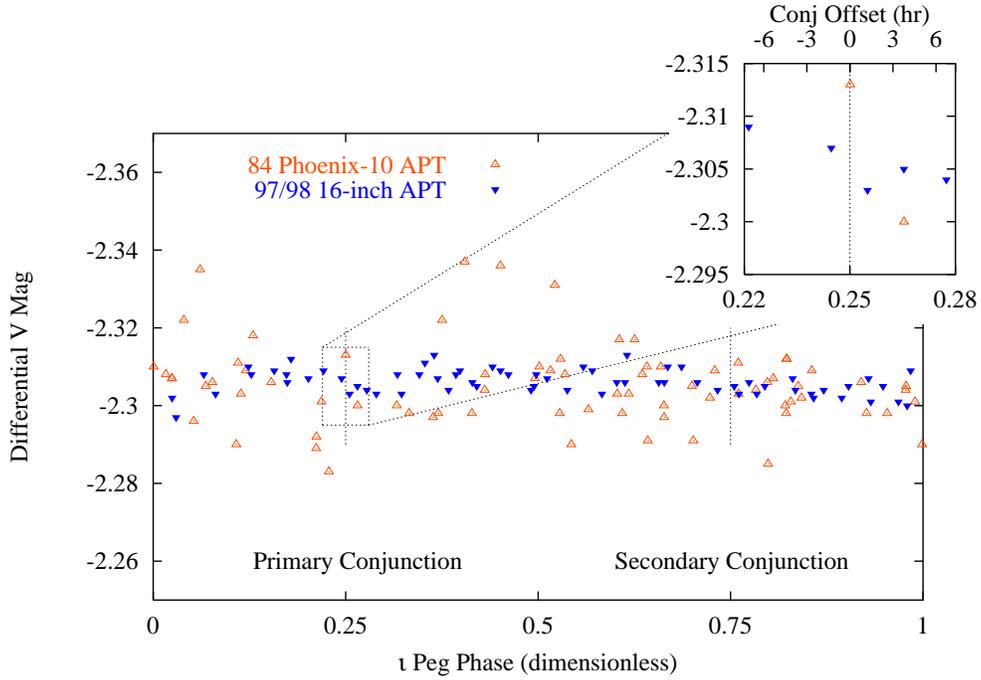}
\caption{Photometric Observations of $\iota$~Peg.  Differential
photometric observations of $\iota$~Peg from the Phoenix-10 APT (open
triangles) and the Vanderbilt/Tennessee State University 16-inch APT
(filled triangles) plotted against orbital phase of the binary
computed following FT.  Phase 0.25 represents a time of conjunction
with the secondary in front (primary eclipse opportunity).  Inset we
show a closeup of the data around the primary eclipse opportunity.
(We have added a second horizontal scale relative to the eclipse
opportunity in units of hours; a full eclipse in the
$\iota$~Peg system would be roughly 7 hours in duration.)  The
photometric observations exclude the possibility of all but the
briefest of grazing eclipses in the $\iota$~Peg system.
\label{fig:iPg_phot}}
\end{figure}

The photometric observations summarized in Table \ref{tab:photometry}
are plotted in Figure \ref{fig:iPg_phot} against orbital phase of the
binary computed from the FT-defined T$_0$ and period.  For
inclinations allowing eclipses of the two components, the phases of
conjunction coinciding with primary and secondary eclipse
opportunities are 0.25 and 0.75 respectively.  FT estimated the total
duration of a central eclipse ($i$ = 90$^\circ$) to be roughly 6.8
hours or 0.027 phase units.  Our photometric observations exclude this
possibility and show no evidence for any partial eclipse to a
precision of around 0.003 mag.  The time of conjunction is uncertain
by no more than a few minutes, and gaps in the data around the time of
conjunction are no larger than about 0.005 phase units (1.2 hours).
Thus, the possibility of all but the briefest of grazing eclipses are
excluded by the APT photometry.  In particular, using the two points
nearest the primary conjunction opportunity (at -1.29 and +1.22 hours
relative to the predicted conjunction respectively) constrain $|90-i|$
to be greater than 4.07$^\circ$ and 4.10$^\circ$ respectively at
greater than 99\% confidence, based on the model diameters and M$_v$
estimates of 3.4 and 5.8 for the primary and secondary components
respectively.

The components of most close binaries with orbital periods less than
about one month rotate synchronously with the orbital period due to
tidal action between the components (e.g.~\cite{Fekel89}).  Such
synchronous rotation is expected in $\iota$~Peg and is confirmed by
the rotational broadening measurements of FT and Gray (1984)
(c.f.~\cite{Wolff97}).  If the G8V secondary, which is much more
convective than the F5V primary, is rotating synchronously, it would
be expected to be photometrically variable on the orbital period at
the level of a few percent due to starspot activity (\cite{Henry99}).
In fact, $\iota$~Peg is listed as a suspected variable star by Petit
(1990), who reports variability at the 0.02 mag level in V.  FT
estimate that the secondary is roughly 2.7 mag fainter in the V band
than the primary, so any apparent photometric variability of the
secondary component will be diluted by a factor of about 12 by the
primary component.

In order to search for this possible photometric variability in
$\iota$~Peg, we performed a periodogram analysis of the 16-inch APT
data.  The analysis reveals a photometric period that is identical,
within its uncertainty, to the spectroscopic period, a result that is
consistent with the assumption of synchronous rotation.  Likewise, the
amplitude of 0.0037 mag, scaled by a factor of 12, results in a 4.4\%
variation, similar to the variability expected from rotational
modulation of the spotted surface of the secondary diluted by the
emission of the primary.  Based on these results, we conclude that
$\iota$~Peg is a low-amplitude variable star.


\section{Summary}

We have presented the visual orbit for the double-lined binary system
$\iota$~Pegasi, and derived the physical parameters of the system by
combining it with the earlier spectroscopic orbit of Fekel and Tomkin.
The derived physical parameters of the two young stars in $\iota$~Peg
are in reasonable agreement with the results of other studies of the
system, and theoretical expectations for stars of these types.  Noted
by FT, the $\iota$~Peg system is nearly eclipsing; because our model
visual orbit is so close to producing observable eclipses we have
further presented high-precision photometric data which is consistent
with our visual orbit model.

$\iota$~Peg represents a prototype of the binary system that PTI is
well-suited to measure; the large magnitude difference between
components in the visible is significantly mitigated in the
near-infrared, making the accurate determination of the system
parameters feasible.

\acknowledgements Part of the work described in this paper was
performed at the Jet Propulsion Laboratory, California Institute of
Technology under contract with the National Aeronautics and Space
Administration.  Interferometer data was obtained at the Palomar
Observatory using the NASA Palomar Testbed Interferometer, supported
by NASA contracts to the Jet Propulsion Laboratory.

Automated astronomy at TSU has been supported for several years by the
National Aeronautics and Space Administration and by the National Science
Foundation, most recently through NASA grants NCC2-977 and NCC5-228 (which
supports TSU's Center for Automated Space Science) and NSF grants HRD-9550561
and HRD-9706268 (which supports TSU's Center for Systems Science Research).

We wish to thank the anonymous referee for his many positive
contributions to the accuracy and quality of this manuscript, and his
forbearance in the review process.

This research has made use of the Simbad database, operated at CDS,
Strasbourg, France.

\end{document}